\documentclass[aps,prl.preprint,superscriptaddress]{revtex4}
\usepackage{graphicx}
\usepackage{longtable,array}
\usepackage{amsmath}
\usepackage{color}
\usepackage{float}
\usepackage{hyperref}
\usepackage{wasysym}
\usepackage{url}
\usepackage{dcolumn}
\usepackage{bm}

\newcommand{\ai}{{\it ab initio}}

\newcommand{\cm}{cm$^{-1}$}

\draft
\begin{document}
\title{Sub-promille measurements and calculations of CO (3--0) overtone line intensities }
\author{Katarzyna Bielska}
\affiliation{Institute of Physics, Faculty of Physics, Astronomy and Informatics, Nicolaus Copernicus University in Toru\'n, Grudziadzka 5, 87-100 Torun, Poland}
\author{Aleksandra A. Kyuberis}
\affiliation{Van Swinderen Institute for Particle Physics and Gravity, University of Groningen, Nijenborgh 4, 9747AG Groningen, The Netherlands}
\author{Zachary D. Reed}
\affiliation{Chemical Sciences Division, National Institute of Standards and Technology,Gaithersburg, Maryland 20899 U.S.A.}
\author{Gang Li}
\affiliation{PTB (Physikalisch-Technische Bundesanstalt), Bundesallee 100, 38116 Braunschweig, Germany}

\author{Agata Cygan}
\affiliation{Institute of Physics, Faculty of Physics, Astronomy and Informatics, Nicolaus Copernicus University in Toru\'n, Grudziadzka 5, 87-100 Torun, Poland}
\author{Roman Ciury{\l}o}
\affiliation{Institute of Physics, Faculty of Physics, Astronomy and Informatics, Nicolaus Copernicus University in Toru\'n, Grudziadzka 5, 87-100 Torun, Poland}

\author{Erin M. Adkins}
\affiliation{Chemical Sciences Division, National Institute of Standards and Technology,Gaithersburg, Maryland 20899 U.S.A.}
\author{Lorenzo Lodi}

\affiliation{Department of Physics and Astronomy, University College London,
Gower Street, London WC1E 6BT, United Kingdom}

\author{Nikolay F. Zobov}
\affiliation{Department of Physics and Astronomy, University College London,
Gower Street, London WC1E 6BT, United Kingdom}

\author{Volker Ebert}
\affiliation{PTB (Physikalisch-Technische Bundesanstalt), Bundesallee 100, 38116 Braunschweig, Germany}

\author{Daniel Lisak}
\affiliation{Institute of Physics, Faculty of Physics, Astronomy and Informatics, Nicolaus Copernicus University in Toru\'n, Grudziadzka 5, 87-100 Torun, Poland}

\author{Joseph T. Hodges}
\affiliation{Chemical Sciences Division, National Institute of Standards and Technology,Gaithersburg, Maryland 20899 U.S.A.}

\author{Jonathan Tennyson}
\email{j.tennyson@ucl.ac.uk.}
\affiliation{Department of Physics and Astronomy, University College London,
Gower Street, London WC1E 6BT, United Kingdom}
\author{Oleg L. Polyansky}
\affiliation{Department of Physics and Astronomy, University College London,
Gower Street, London WC1E 6BT, United Kingdom}

\

\date{\today}


\begin{abstract}
Intensities of lines in the near-infrared second overtone  band (3--0) of $^{12}$C$^{16}$O are measured and calculated to an unprecedented degree of precision and accuracy. Agreement between theory and experiment to better than 1 $\permil$ is demonstrated by results from two laboratories involving two independent absorption- and dispersion-based cavity-enhanced techniques.  Similarly, independent Fourier transform spectroscopy measurements of stronger lines in this band yield mutual agreement and consistency with theory at the 1 $\permil$ level. This set of highly accurate intensities can provide an intrinsic reference for reducing biases in future measurements of spectroscopic peak areas.
\end{abstract}
\maketitle

Spectroscopic measurements of transition frequencies (line positions) of gaseous molecules provide some of the most accurate measurements in the whole of science, with relative uncertainties as low as a few parts in $10^{12}$ \cite{Reed:20}. Even  standard laboratory set-ups can routinely provide line positions of rotation-vibrational lines in the infrared (IR), microwave and optical region with relative uncertainties of a few parts in $10^{8}$. 
The situation is very different for line intensities, for which the level of 
accuracy achievable by both experiments and theory is usually much lower, 
typically in the range 1--20 \%. Nevertheless, over the past twenty years or so 
it has become possible in some cases to obtain line intensities with a relative 
standard uncertainty less than 1 \% \cite{Lisak2009}.  More recently intensity 
measurements with combined uncertainties at the pro-mille (1 \permil) level have 
been realized  \cite{Cygan2019,Fleisher2019,D_LongGRL2020,Birk_2021}.  Such an 
accuracy is required for several applications, including detailed observations 
of the various constituents of the Earth's atmosphere, as well as analyses of 
the atmospheres of  celestial bodies such as other Solar System planets, 
exoplanets and cool dwarf stars.  Highly accurate line intensities might also 
become useful for metrological purposes, for example for new standards of 
temperature \cite{Arroyo93}, and pressure \cite{Wehr03,20Gaiser.CO2} although 
the accuracy requirements are one to two orders of magnitude higher than 
presently possible.  Nevertheless, absolute optical measurements of isotopic 
composition based on line intensities having uncertainties at the promille level 
\cite{21NaturePhys.CO2} have recently been shown to be competitive with 
traditional approaches based on  high-precision isotope-ratio mass-spectrometry  
of reference materials.  

Agreement between experiment and theory for some CO$_2$ lines in the near-IR and 
IR regions to a level better than 3 \permil\ was reported by some of us 
\cite{jt613}. Key to the success of this  work was collaboration between 
experiment and theory which allowed the \ai\ theoretical model to provide CO$_2$ 
transition intensities for atmospheric studies \cite{jt691s}.

This paper constitutes a contribution towards the ambitious goal of reliably 
measuring and computing molecular line intensities to sub-promille accuracy.
To this end, we turn to the calculation and measurement of intensities in the 
(3--0) band of $^{12}$C$^{16}$O. Compared to CO$_2$, CO is more amenable to 
accurate theoretical calculations and far less susceptible to experimental 
complications caused by adsorption and desorption from the walls of sample 
cells. There are also fewer overlapping spectra from adjacent lines including 
those of its other isotopologues. To reduce statistical and systematic 
measurement uncertainties, we included independently measured line intensities 
from three laboratories, in which each group had metrology-grade expertise in 
quantitative spectroscopy with traceability to the Syst\`eme International (SI). 
This set of experiments involved different measurement techniques and different 
gas samples for each group.
Except for one value from Ref. \cite{Cygan2019}, measurements were arranged in a 
blind approach so that experimenters did not know other's results before the 
final line intensity data were revealed. After combining results from all 
laboratories and substantially different measurement techniques described below, 
the dynamic range of the line intensities is $\approx$1400:1.  In the remainder 
of the article all reported intensities are compared to the present \ai\ 
theoretical calculations and to literature values.

Seven lines namely R23, R26, R27, R28, R29, P27 and P30, were independently 
measured at the Nicolaus Copernicus University (NCU) and at the National 
Institute of Standards and Technology (NIST) using two different laser-based 
techniques. Measurements at NCU were made using the recently developed 
cavity-mode dispersion spectroscopy (CMDS) technique \cite{Cygan2015}, which 
involves mode-by-mode measurements of shifts (dispersion) in cavity mode 
frequencies under steady state laser excitation for gas pressures between 
0.4~kPa and 13.1~kPa.
Spectra at NIST were acquired using the comb-linked cavity ring-down 
spectroscopy (CRDS) technique \cite{Lin2015,Reed2020}. 
These CRDS spectra were acquired using mode-by-mode measurements of the 
laser-pumped ring-down cavity decay rate for gas pressures between 8.7~kPa and 
26.6~kPa. Importantly, we note that unlike CMDS, CRDS measures light absorption 
by the sample instead of sample dispersion. Both of these observables are 
proportional to line intensity and connected by the principle of causality 
through the real and imaginary components of the complex-valued resonant 
susceptibility of the absorber \cite{Lehmann99}.

To provide more extensive coverage of the (3--0) CO band, another set of  
measurements for stronger lines (P22 to R22) in the same band was measured at a 
gas pressure of 10.2 kPa at PTB \cite{Werwein17.N2O} using the Fourier-transform 
spectroscopy (FTS) technique. The implementation of FTS here involves an 
incoherent light source coupled into a scanning inteferometer to produce an 
interferogram, from which the spectrum is obtained through Fourier 
transformation.

 All lines were fit with the Hartmann-Tran profile (HTP) \cite{Ngo2013} or 
limiting cases of this profile. These analyses included first-order line-mixing 
effects, and both NCU and NIST used multispectrum fitting approaches 
\cite{Pine2001}. Full details regarding the experiments and theoretical analysis 
are available in the Suplemental Material (SM) which includes Refs. 
\cite{Truong2013,Hendricks2010,Konefal2020,jt859,jt609,02PeDuxx.ai,jt509,Hartmann86.CO2, Devi18.CO2, Zou02.CO2, Varanasi75.CO2, Sung05.CO2, Regalia05.CO2}; 
the SM also provides  sample figures illustrating the spectra. Each group independently 
identified and evaluated both systematic and statistical sources of measurement 
uncertainty.  In summary, the maximum systematic uncertainty components among 
all experiments included pressure (0.7 \permil), temperature dependence of 
sample density and line intensity (0.9 \permil), sample isotopic composition and 
purity (0.4 \permil), and spectral modeling (1 \permil). The maximum component 
of the uncertainty driven by statistical variations in the measurements was 1.4 
\permil.  For the NIST measurements, the effect of digitizer nonlinearity was 
also included (0.2 \permil), whereas for the FTS measurements at PTB uncertainty 
in sample pathlength was 0.12 \permil.  Adding all components in quadrature 
resulted in line- and institution-dependent relative combined standard 
uncertainties ranging from 1 $\permil$ to 1.2 $\permil$ for NCU, 0.9 $\permil$ 
to 1.8 $\permil$ for NIST, and an average value of 1.3 $\permil$ for PTB.

The accuracy of intensity calculations (both purely \ai\ and semi-empirical) is 
determined by the accuracy of the wavefunctions and dipole moment curve (DMC), 
where uncertainty of the wavefunctions is based on the potential energy curve 
(PEC) and solution of the nuclear-motion Schr\"odinger equation. Here we used an 
empirical PEC from Coxon {\it et al.} \cite{04Coxon.CO2}, which reproduces the 
CO transition frequencies within experimental uncertainty. DMCs were computed 
\ai\ using the electronic structure package MOLPRO\cite{Molpro:JCP:2020} at the 
multi-reference configuration interaction (MRCI) level of theory with a Davidson 
correction (+Q) and a relativistic correction using an aug-cc-pCV6Z  basis set.

\begin{figure}
\includegraphics[width=0.45\textwidth]{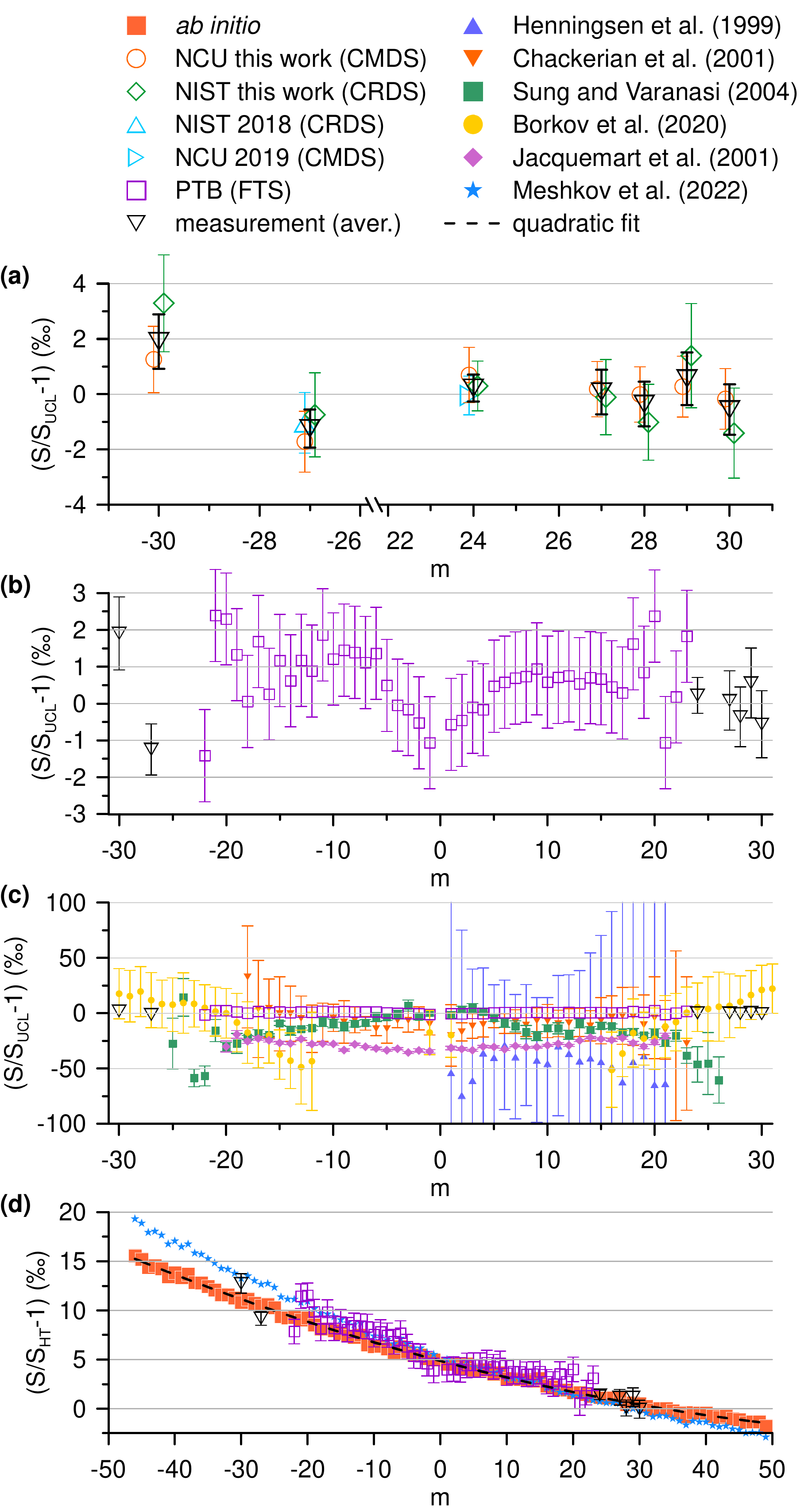}
\caption{Summary of present experimental and theoretical results for the (3--0) band of $^{12}$C$^{16}$O, and comparison to literature data.
Panel (a): Comparison of NCU and NIST intensity measurements, $S$, for high- $J$ lines, expressed as  $(S/S_{UCL}-1)$, where $S_{UCL}$ are the present theoretical intensities. The black triangles represent weighted averages of the measurements. NCU 2019 indicates the measurement from Ref. \cite{Cygan2019} and NIST 2018 indicates a prior measurement conducted at NIST in 2018. Note the break in the $x$ axis.
Panel (b): Analogous comparison of low-$J$ PTB intensity values and average high-$J$ values from panel (a).
Panel (c): $m$ dependence of $(S/S_{UCL}-1)$ where $S$ corresponds to literature data \cite{99HeSuMo,01JaMaDa,01ChFrGiv,04SuVa,20BoSoSo} (closed symbols) and present measurements (open symbols).  Note the extended range of the  $y$-axis scale compared to the other panels.
Panel (d): $m$ dependence of $(S/S_{HT}-1)$ where $S$ corresponds to experimental (open symbols) and theoretical (orange squares) intensities from this work as well as semi-empirical intensities from Ref. \cite{Meshkov2022} (blue stars). $S_{HT}$ are values from HITRAN 2020 \cite{jt836s} intensities scaled to 100 \% $^{12}$C$^{16}$O. The dashed line is a second-order polynomial fit of the ratio $(S_{UCL}/S_{HT}-1)$ versus $m$.  This function, $f(m) = a_0 + a_1m + a_1m^2$ provides a convenient mapping of the HITRAN 2020 intensities on to the present theoretical values. Here, $a_0$ = 4.865 $\permil$, $a_1$ = -0.1789 $\permil$, and $a_2$ = 0.001020 $\permil$.  Note the differing $x$ axes for the various panels.
}
\label{fig:obs.-calc.3-0}
\end{figure}


Figure \ref{fig:obs.-calc.3-0} summarizes the present results in four panels which show relative differences in intensity, $S$, versus, $m$, where $m = -J$ ($P$-branch), $m=J+1$ ($R$-branch), and $J$ is the lower-state rotational quantum number.  In Fig. \ref{fig:obs.-calc.3-0}a we present the high-$J$ line intensities measured by NCU and NIST,  relative to the present theoretical calculations.  The standard deviation of the relative differences between the experimental and calculated intensities as well as the relative differences between the measured quantities themselves are around 1 $\permil$ (excluding the relatively weak P30 transition), see Table \ref{tab:exp}.   To consolidate the measured intensities, we also evaluated the weighted mean intensity, $S_{av}$, at each value of $m$.  The relative difference, $(S_{av}/S_{UCL} -1)$, averaged over the remaining six lines  (see black triangles in Fig. 1a) gives a mean value of -0.2 $\permil$ with a standard deviation of 0.7 $\permil$. These results are consistent with the uncertainties in the NCU and NIST measurements.

Special attention should also be paid to the R23 line intensity for which three independent sets of measurements  (two at NCU and one at NIST) were made. This line was first measured by the NCU group with the method described above \cite{Cygan2019} and a second time for the present study. We also note that the CMDS measurement technique at NCU was compared to an alternative technique based on cavity buildup dispersion spectroscopy \cite{Cygan2021}. In that experiment, the R23 peak area was measured by both techniques and yielded a difference of 0.3 $\permil$, which is consistent with the combined uncertainties in both measurements of spectroscopic area.  In the present results, the weighted average of the experimental intensities for the R23 line differs from the theoretical value by 0.2 $\permil$, with an uncertainty of 0.5 $\permil$. Given that the theoretical calculations and measured intensities of the most accurately measured lines agree to within the estimated uncertainties, the demonstrated 0.5 $\permil$ agreement is unlikely to be fortuitous.

The FTS results from PTB for the stronger lines are given in Fig.~\ref{fig:obs.-calc.3-0}b and a representative subset of these data are given in Table \ref{tab:exp}, see SM for all data. In summary, the set of relative differences between the intensities measured by PTB and the theoretical predictions has a mean value of 0.7 $\permil$ with a standard deviation of 0.9 $\permil$. For the strongest 50 \% of the lines, the mean value is 0.2 $\permil$ and the standard deviation is 0.4 $\permil$, thus quantifying excellent agreement with the theory. Also shown in this figure are the weighted average data from the laser based measurements from NCU and NIST. Upon comparison of all measured line intensities, there is no evidence of substantial (much greater than 1 $\permil$) systematic deviation about the reference UCL values.  This demonstrated consistency in measured intensities puts an upper bound on any unaccounted-for systematic bias in the three sets of measurements performed using the three different experimental techniques considered here.

\begin{table*}[!ht]
\caption{Comparison of experimental (NCU, NIST and PTB) and calculated line intensities in the (3--0) of CO. Reported intensities, $S$, are based on the reference temperature $T$ = 296 K, and scaled to 100 \% relative abundance of the $^{12}$C$^{16}$O isotopologue. Intensity units are in cm$^{2}$ cm$^{-1}$/molecule. Line wavenumbers are from HITRAN 2020 \cite{jt836s}.}
\label{tab:exp}
\begin{tabular}{cccccccc}
\hline\hline
line & wavenumber	&$S_{\textrm{NIST}}$ & $S_{\textrm{NCU}}$ &$S_{\textrm{PTB}}$ & $S_{\textrm{NIST}}/S_{\textrm{UCL}}-1\quad$ &$S_{\textrm{NCU}}/S_{\textrm{UCL}}-1\quad$ & $S_{\textrm{PTB}}/S_{\textrm{UCL}}-1$    \\
     &  \cm     &   &   &  & $\permil$ & $\permil$ & $\permil$   \\
\hline
P30 & 6190.07&1.5790(28)E-26& 1.5758(18)E-26 &   &3.3&1.3&   \\
P27 & 6210.25&7.3680(120)E-26& 7.3608(76)E-26&   &-0.8&-1.7&   \\
P27 & 6210.25& 7.3658(82)E-26$^a$  & &&-1.0& \\
P20 & 6253.78 &&& 1.3514(18)E-24 &&& 2.3 \\
P15 & 6281.82 &&&5.7460(75)E-24 &&& 1.2 \\
P10 & 6307.29 &&& 1.3675(18)E-23 &&& 1.2 \\
P5 & 6330.17 &&& 1.5321(20)E-23 &&& 0.5 \\
R5 & 6371.30 &&& 2.1075(28)E-23 &&& 0.6 \\
R10 & 6385.77 &&& 1.9505(26)E-23 &&& 0.7 \\
R15 & 6397.58 &&& 8.989(12)E-24 &&& 0.5 \\
R20 & 6406.70 &&& 2.3469(31)E-24 &&& -1.1 \\
R23&6410.88&8.1687(74)E-25& 8.1719(77)E-25 &  &0.3&0.7&  \\
R23&6410.88& & 8.1659(58)E-25$^b$ & &&-0.1&\\
R26&6414.08&2.3661(33)E-25& 2.3668(24)E-25 &  &-0.1&0.2&   \\
R27&6414.93&1.5034(21)E-25& 1.5049(16)E-25 &  &-1.0&0.0&   \\
R28&6415.67&9.397(18)E-26& 9.386(11)E-26&   &1.4&0.3&   \\
R29&6416.30&5.7300(94)E-26& 5.7371(63)E-26&  &-1.4&-0.2&  \\

\hline

\multicolumn{6}{l}{$^a$Measurement conducted at NIST in 2018.}\\
\multicolumn{6}{l}{$^b$Measurement conducted at NCU in 2019 \cite{Cygan2019}.}

\end{tabular}
\end{table*}

We present both literature (closed symbols) intensities and our experimental values (open symbols) relative to our theoretical values in Fig. \ref{fig:obs.-calc.3-0}c. In contrast to the other panels in Fig. \ref{fig:obs.-calc.3-0}, note the significantly extended range along the $y$-axis. The scatter and the $m$-dependent structure of the literature data about the theoretical intensities are more than an order of magnitude greater than those of our measured values, which are nearly indistinguishable from zero on the chosen scale.  We reemphasize that the mutual agreement with theory of our laser-based and FTS spectroscopic line intensity measurements (both of which are SI traceable and can be considered as primary measurements of line intensity) is unprecedented. These measurements span a sufficiently broad range of rotational quanta to confirm both the calculated $J$-dependence (band shape) of the component intensities as well as the total band intensity (given below). Combining the present results from all three experimental techniques and all lines yields an average deviation between experiment and theory of 0.6 $\permil$ with a standard deviation of 0.9 $\permil$ - representing a more than order-of-magnitude improvement in measurement precision and accuracy compared to the literature values given in Fig. \ref{fig:obs.-calc.3-0}c.
We note that with the exception of Ref. \cite{99HeSuMo}, all the literature data  presented were acquired with the FTS technique. The relatively high precision and accuracy achieved with the present FTS measurements is ascribed to several factors including precise temperature stabilization, characterization of the sample path length and instrument line shape function, single-polynomial fits to the baseline, high signal-to-noise ratio (nominally 2000:1), and the use of an InGaAs detector with high linearity.

In Fig. \ref{fig:obs.-calc.3-0}d we also compare the present experimental and theoretical results to the HITRAN 2020 \cite{jt836s} intensities for the (3--0) band of $^{12}$C$^{16}$O.  The latter intensities are based on a semi-empirical dipole-moment function which is determined by global fitting to measured intensities from multiple vibrational bands of this molecule. Comparison of our theoretical intensities (see Fig. 1d and SM) with those from HITRAN 2020 reveals a nearly quadratic trend  with rotational quantum number. This quantity is about 5 $\permil$ near band center  and exhibits a slope of nominally -0.2 $\permil$. In addition to this discrepancy in the rotational dependence of line intensity, the HITRAN 2020 band intensity (based on summing over lines P46 to R48) is about 4.6 $\permil$ smaller than our theoretical value of $4.7361 \times 10^{-22}$ cm$^{2}$ cm$^{-1}$/molecule.

We also compare our results to a semi-empirical line list for CO (see Fig. \ref{fig:obs.-calc.3-0}d; blue stars) which has recently been created. Similar to generation of the HITRAN 2020 line list, the new one was based on a global fit  of a parameterized DMC to multiple CO vibrational bands (up to the fifth overtone) \cite{Meshkov2022}. Comparison to our results shows good agreement (at the pro-mille level) both with the present experimental data and \ai\ calculations. However, the two sets of calculated values tend to diverge with increasing $J$ in the $P$-branch, with differences of 2.0 $\permil$ at $m$ = -25 that increase to 3.7 $\permil$ at $m$ = -46.  Thus, our comparison between the present \ai\ intensities and the semi-empirical results from \cite{Meshkov2022} also provides a measure of uncertainty in the latter values, which without the present results cannot be easily assessed.  This difficulty arises because uncertainties in the intensity data from prior measurements, which were used in the global fit of \cite{Meshkov2022} are often underestimated as can be seen in Fig. \ref{fig:obs.-calc.3-0}c. For example, it would have been difficult identify the systematic offset in the $P$-branch (at the promille-level) without reference to the present results.

Our theoretical model should reliably predict intensities for all vibrational bands of CO. Unfortunately, experimental results with sufficient precision and accuracy (such as those demonstrated here for the (3--0) band) needed to validate this predictive capability in the other CO bands at the sub-promille level do not exist. Notwithstanding this limitation, we can say that for the (1–-0), (2--0), and (4--0) bands, our new model for CO intensities is consistent with other experimental results in the literature within their reported percent-level uncertainties (see SM) \cite{15GangLi.CO2,Hartmann86.CO2, Devi18.CO2, Zou02.CO2, Varanasi75.CO2, Sung05.CO2, Regalia05.CO2,Devi12.CO2}. Our theoretical line intensities for these bands are given in Tables II-V of the SM. Higher overtones, including the (5--0) band and beyond, will require additional work because the calculated intensities are affected not only by the accuracy of the quantum chemistry calculations, but also by that of the DMC functional form \cite{Medvedev21.CO2}. 

We propose that the present theoretical (3--0) CO line intensities could be used as intrinsic spectroscopic references to improve measurement accuracy in the case of techniques such as FTS and cavity-enhanced spectroscopy (CEAS) \cite{Gianfrani:99} which require knowledge of the optical path length. (See SM for details).  For cases of comparable optical thickness and line shape in both the reference and unknown spectra, this approach could help reduce biases in retrieved absorber number density that depend on signal-to-noise ratio and are driven by non-linear dependence of the spectrum on absorber number density. This approach would be nearly cost-free and has the potential to reduce systematic relative uncertainties towards promille levels in the case of long-path spectrometers.

We close this article by noting that the present work is the foundation for an emerging international effort by the Consultative Committee for Amount of Substance (CCQM).  This body meets regularly at the International Bureau of Weights and Measures (BIPM) in Sèvres, France and informs the International Committee for Weights and Measures (CIPM) charged with promoting world-wide uniformity in the SI of measurement units.  A new Task Group on Advanced Spectroscopy (TGAS) within CCQM has recently been initiated, which comprises gas metrology experts and spectroscopists from several National Metrology Institutes and other technical communities who are developing various laser-based techniques and traditional methods using Fourier-transform spectroscopy.  The purpose of this task group is to promote the development, realization and harmonization of these primary spectroscopic methods for amount of substance through rigorous intercomparison experiments.  This effort will leverage expertise in gas mixture preparation, high-resolution spectroscopy, and quantum-chemistry calculations to enable robust, SI-based uncertainty budgets for spectroscopic gas analysis.


\section*{Acknowledgments}
The reported study was funded by RFBR according to the research project No. 18-02-00705 and 18-32-00698.
he research conducted at NCU was supported by National Science Centre, Poland project no. 2018/30/E/ST2/00864, 2015/18/E/ST2/00585 and 2020/39/B/ST2/00719, and it was part of the program of the National Laboratory FAMO in Toruń, Poland.
The work at NIST was supported by the NIST Greenhouse Gas and Climate Science Measurements Program.
NFZ and OLP acknowledge support by State Project IAP RAS No. 0030-2021-0016.
GL acknowledges technical support from Alexandra Domanskaya and Kai-Oliver Krauß. O.L.P. acknowledges support from the Quantum Pascal project 
18SIB04, which received funding from the EMPIR programme co-financed by the 
Participating States and the European Union’s Horizon 2020 research and innovations 
programme. OLP is grateful to Nino Ipfran for technical support. 






\newpage
\section*{Supporting Material}
\begin{center}
{Sub-promille measurements and calculations of CO (3--0) overtone line intensities}\\
{Katarzyna Bielska, Aleksandra A. Kyuberis, Zachary D. Reed, Gang Li,  Agata Cygan, Roman Ciury{\l}o, Erin M. Adkins,
 Lorenzo Lodi, Nikolay F. Zobov, Volker Ebert, Daniel Lisak, Joseph T. Hodges, Jonathan Tennyson and Oleg L. Polyansky}
\end{center}

\section{Measurements}

\subsection{Experiment  at NCU} \label{sec:torun}
Measurements at NCU  were made using the recently developed cavity-mode dispersion spectroscopy (CMDS) technique \cite{Cygan2015}, which is based on the determination of dispersive shifts in the modes of a high-finesse optical cavity that are induced by molecular absorption.  Therefore both axes of measured spectrum are retrieved from measurements of optical frequency shifts. We have shown that CMDS provides high-accuracy determinations of line intensity, has a wide dynamic range, and is highly immune to nonlinearity of the detection system \cite{Cygan2019}. The spectrometer itself has been described in detail in the previous work \cite{Cygan2019}. 
The optical cavity consists of two spherical mirrors of nominal intensity reflectivity $R$ = 0.999 923 at the wavelength, $\lambda$ = 1.6~$\mu$m corresponding to the transitions under investigation and $R=$  0.98 at $\lambda$= 1.064~$\mu$m corresponding to the frequency-stabilized Nd:YAG laser which serves as a frequency reference for the stabilization of the cavity length. 
This leads to the relative stability of the cavity resonances frequencies below $3\times10^{-11}$.
The probe laser, which is an external cavity diode laser (ECDL), is frequency-locked and spectrally narrowed with the Pound-Drever-Hall technique to a selected cavity resonance that is detuned from the transition frequency by a few GHz. An orthogonally polarized beam from the same laser is phase-modulated at radiofrequencies with a broadband electro-optic modulator (EOM) to produce tunable sidebands, one of which is used to probe consecutive cavity modes. These steady state transmission spectra provide mode positions from which sample-induced dispersion can be determined. 
The mode positions are measured relative to the selected mode frequency, therefore the relative accuracy of $3\times10^{-11}$ leads to sub-Hz  accuracy of the local frequency axis for detunings below 20 GHz, used in our measurements.
The cavity temperature was actively stabilized to 296.00~K with a combined standard uncertainty of 30~mK.
Sample pressure was determined with a calibrated capacitance diaphragm manometer (MKS Baratron 690A) with a relative combined standard uncertainty of 0.05 \%. Measurements were done with a commercial sample of CO (0.999 97 purity) produced by the reaction of water with petrogenic natural gas having an estimated $\delta^{13}$C$_{\textrm{VPDB}}$ content of -40 $\permil$.

\begin{figure}
\includegraphics[width=0.45\textwidth]{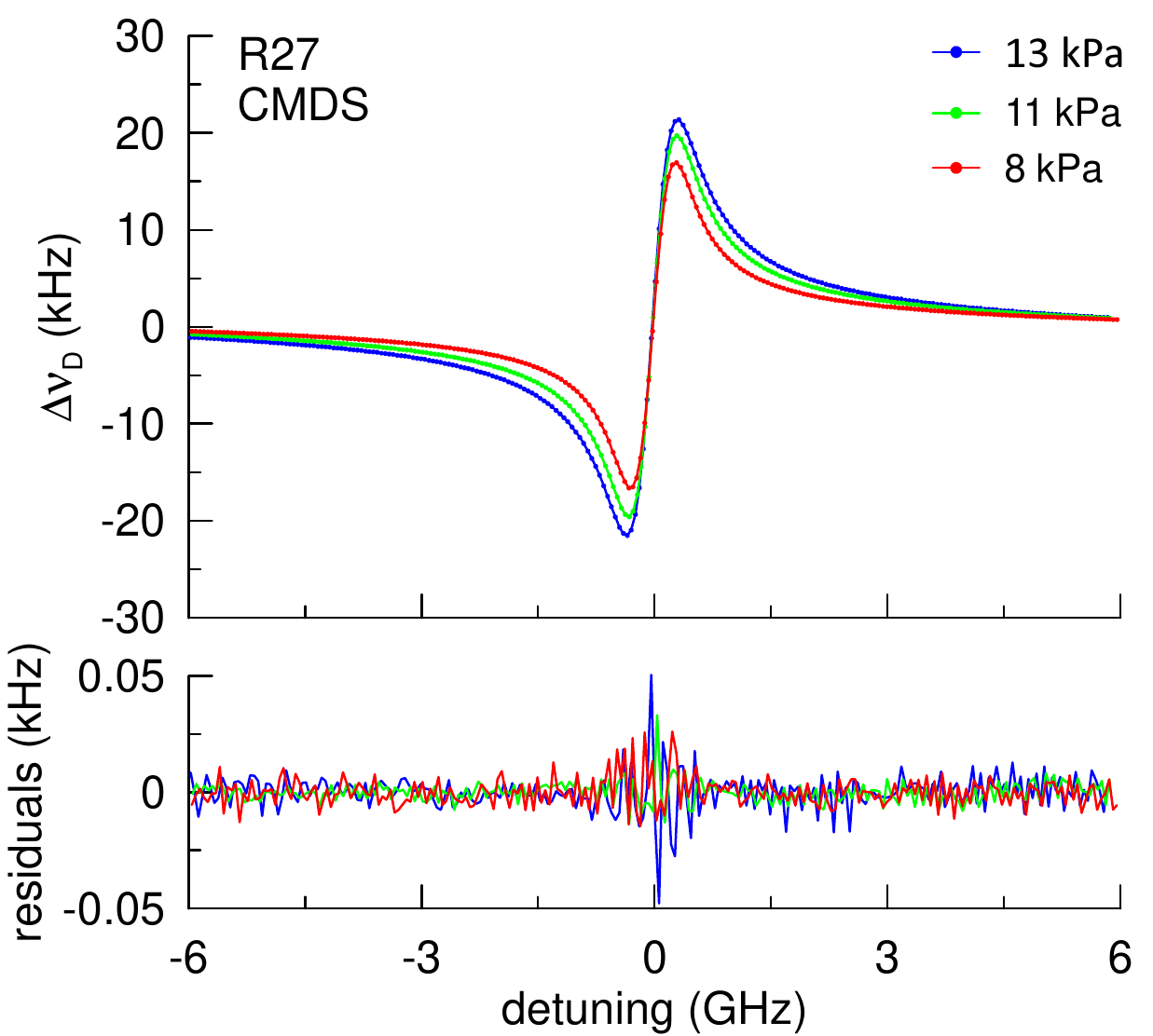}
\caption{Top panel: Self-perturbed line R27 measured at NCU with the CMDS technique at nominal pressures of 8~kPa, 11~kPa and 13~kPa and temperature of 296~K. Measured points are shown with dots, whereas lines indicate fitted profiles.
Bottom panel: Fit residuals (observed - calculated) obtained from a multispectrum fit of the HTP (with first-order line-mixing) to the measured spectra. The ratio of the maximum peak-to-peak mode shift to the root-mean-square of the fit residuals for the highest-pressure spectrum is $5400$. }
\label{fig:ncu}
\end{figure}

Spectra were acquired in a range of pressures from 0.4 kPa to 3.3 kPa for the line R23 and up to 13.1 kPa for other transitions. The achieved signal-to-noise ratio was between 1000:1 and 8000:1.  They were fit with the Hartmann-Tran profile (HTP) \cite{Ngo2013} using multispectrum fit approach. As an example, spectra measured for the R27 line together with the fit residuals are shown in Fig. \ref{fig:ncu}. The fitting parameter $\eta$, accounting for the correlation between phase- and velocity-changing collisions, has negligible effect on the line intensity. In spite of the low measurement pressure range, we observed the presence of the line-mixing effect, which was added to the HTP. 
Inclusion or exclusion of the above two effects changes determined line intensities by less than 0.1$\permil$.
Contributions to the relative systematic uncertainty include pressure (0.5 $\permil$) and temperature (between 0.3 $\permil$ and 0.7 $\permil$) measurement, spectrum modeling (0.6 $\permil$), sample isotopic composition (0.4 $\permil$), and sample purity (0.03 $\permil$). The relative statistical uncertainty varied between 0.05 $\permil$ and 0.5$\permil$.

We should note that the effects of scattering, such as Rayleigh or Rayleigh-Brillouin scattering, weakly depend on the probe laser wavelength. Therefore, it causes an approximately constant background in the spectral range of a single molecular line in absorption. This background is fitted out in cavity-enhanced techniques (eg. cavity ring-down spectroscopy) together with other losses, and in particular cavity mirror losses, as a linear background in measured spectra. In the dispersive CMDS technique, this kind of broadband effect causes the change of the cavity free spectral range (FSR), which is also a fitted parameter in the data analysis.

\subsection{Experiment at NIST} \label{sec:nist}
All measurements at NIST were performed using the cavity ring-down spectroscopy (CRDS) method using the spectrometer originally presented in Ref. \cite{Lin2015}.  This cavity comprised two high-reflectivity mirrors, ($R$ = 0.999 976) separated nominally by 140 cm, with an actively stabilized cavity length.  Similar to the approach implemented by the NCU group, the data point labeled NIST 2018 was based on spectra acquired using the frequency-agile rapid scanning (FARS) method \cite{Truong2013} . This laser scanning technique involves the generation of radiofrequency sidebands using an EOM, combined with the selection of a single sideband by using the ring-down cavity as a frequency filter. For the NIST 2018 data, the cavity length was locked to a frequency-stabilized HeNe laser and the spectrum detuning axis was expressed in terms of multiples of the cavity longitudinal mode spacing. All other NIST data reported here involved an upgraded version of the same spectrometer in a configuration referred to as comb-linked (CL)-CRDS as described in Ref. \cite{Reed2020}. In this case, the cavity length was stabilized with respect to an external-cavity diode laser (ECDL) operating from $\lambda$= 1.55 $\mu$m  to 1.63 $\mu$m. The probe laser was in turn frequency locked to a commercial octave-spanning optical-frequency comb (1 $\mu$m to 2 $\mu$m wavelength) synchronized to a Cs clock. As with the first set of NIST data from 2018, the ECDL output was also scanned through successive cavity modes  using the FARS scheme \cite{Truong2013}.  In all cases, following the excitation of each mode, the probe beam was extinguished and cavity ring-down decay signals were measured with an InGaAs photoreceiver followed by a reference-grade 16-bit digitizer (with the exception of the NIST 2018 experiments which used another digitizer whose response was later validated by the reference digitizer). To characterize potential biases in the measured decay times using the reference-grade digitizer, we generated synthetic decay signals with an arbitrary waveform generator and recorded them with the digitizer, revealing that digitizer nonlinearity was less than 0.2 $\permil$ \cite{Fleisher2019}. Also, measurements made with two InGaAs photoreceivers from different manufacturers and with slightly different electronic bandwidths of 700 kHz and 500 kHz, respectively, were found to be statistically indistinguishable. 

Sample gas pressures were measured with a relative uncertainty less than 0.1 $\permil$ using a silicon resonant sensor with SI-traceability to the primary manometer pressure standard at NIST, which incorporates a high-precision ultrasonic interferometric readout method (\cite{Hendricks2010}).  The temperature of the ring-down cavity was measured using a NIST-calibrated platinum-resistance thermometer (20 mK uncertainty) in good thermal contact with the cell walls. The mean temperature was 296.60 K with a long-term stability of $\pm$ 20 mK and a maximum axial temperature difference of 30 mK, to give a combined temperature uncertainty of 40 mK.   Intensities were corrected to 296 K using the mean temperature for each spectrum and the known partition function for $^{12}$C$^{16}$O and corresponding lower-state energy. Propagating the temperature uncertainty into this temperature correction leads to line-dependent uncertainty components ranging from 0.4 $\permil$ to 0.9 $\permil$.

Gas samples introduced into the ring-down spectrometer were provided by two gas cylinder CO/N$_2$ mixtures with NIST-certified mole fractions, $x_{CO}$, of $x_{CO}$ = 0.01016000(55) (CAL7547) and $x_{CO}$ = 0.119860(95) (CAL7563), respectively. These mixtures were prepared by the Gas Standards Metrology Group at NIST using gravimetric methods with traceability to the kilogram and definition of the mole.  As with the sample gas measured by the NCU group, we assume petrogenic origin for the parent gas with an expected $\delta^{13}$C$_{\textrm{VPDB}}$ content of nominally -40 $\permil$. Measured intensities are converted to 100 $\%$ relative abundance of the $^{12}$C$^{16}$O$_2$ (26) isotopologue upon dividing by the HITRAN reference value for this ratio, $\chi_{26}$ = 0.986544. For the present measurements from NIST and NCU, not correcting for the expected $\delta^{13}$C$_{\textrm{VPDB}}$ value leads to an uncertainty component of 0.4 $\permil$ in the reported intensities.

\begin{figure}
\includegraphics[width=0.45\textwidth]{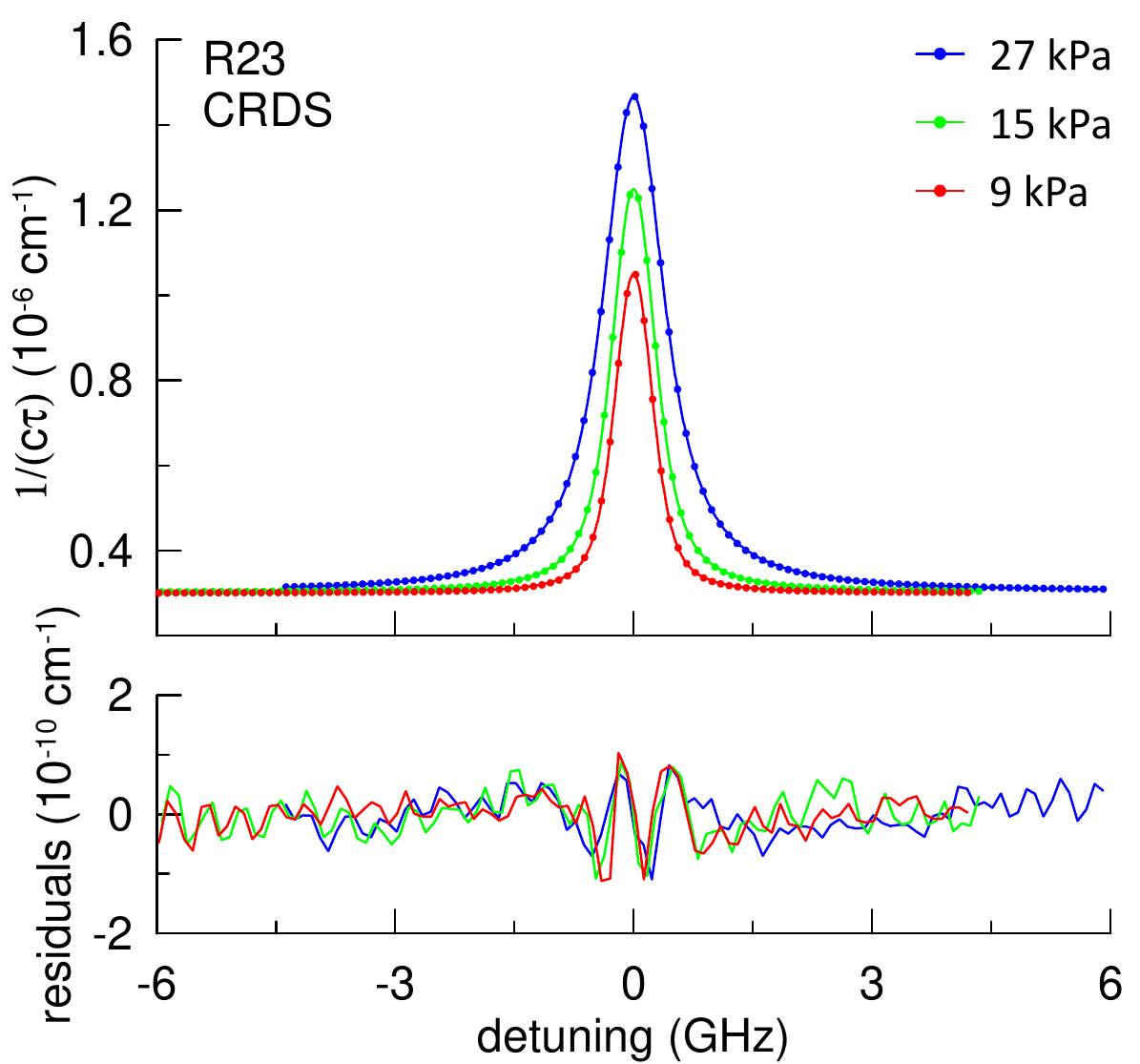}
\caption{Top panel: Three spectra of the R23 line measured at NIST using the CRDS technique. The sample was a CO/N$_2$ mixture with $x_{CO}=0.01$, a temperature of 296~K, and total pressures of 9~kPa, 15~kPa and 27~kPa, respectively, where all specified values are nominal. Measured points are shown with dots, whereas lines indicate fitted profiles.
Bottom panel: Fit residuals (observed - calculated) obtained from individual fits of the HTP (with first-order line-mixing) to each measured spectrum.  The ratio of the maximum absorption to the root-mean-square of the fit residuals for the highest-pressure spectrum is $3.4\times10^4$.}
\label{fig:nist}
\end{figure}

Individual spectra were acquired under static conditions at six pressures over the  range 8.7 kPa to 26.6 kPa and fit with three advanced isolated line profiles plus first-order line mixing. Profiles considered included three variants of the Hartmann-Tran profile (HTP) \cite{Ngo2013}, which included the full HTP, $\beta$-HTP \cite{Konefal2020}, and HTP($\eta$=0) where $\eta$ is the correlation between velocity- and phase-changing collisions.  Spectra were fit both individually and in constrained multi-spectrum analyses, yielding maximum relative differences in intensities of approximately 0.2 $\permil$ for all cases considered. We note that omission of the line mixing model changed the fitted intensities by as much as 1 $\permil$. As an example, spectra measured for the R23 line together with the fit residuals are shown in Fig. \ref{fig:nist}.

Systematic uncertainties considered included measurements of pressure, mean sample gas temperature and gradients, temperature dependence of the correction to line intensity, sample composition, linearity of the digitizer used in recording the ring-down signals, and statistical components corresponding to uncertainty in the fitting measured peak areas and measurement reproducibility.  Quadrature summation of the systematic uncertainties from pressure and temperature ($0.5\ \permil$ to 0.9 $\permil$), spectrum modeling (0.2 $\permil$), sample isotopic composition and purity (0.4 $\permil$), digitizer non-linearity (0.2 $\permil$), with statistical values based on measurement reproducibility and fit uncertainties (0.6 $\permil$ to 1.4 $\permil$) resulted in line-dependent relative combined standard uncertainties between 0.9 $\permil$ (R23) and 1.8 $\permil$ (R28), with an average over all lines of 1.4 $\permil$.

\subsection{Experiment at PTB} \label{sec:ptb}

The absorption spectrum of CO was recorded with a Bruker 125 HR FTIR 
instrument of the EUMETRISPEC facility at PTB. The facility has been previously 
described in detail in the measurement of the N$_2$O overtone band near 2 $\mu$m
region \cite{Werwein17.N2O}. 
In brief, the spectrometer was equipped with a CaF$_2$ beamsplitter, a tungsten lamp, and a room temperature InGaAs detector. 
The spectral resolution was set to 0.012 cm$^{–1}$ and an aperture size of 1.5 mm. The apodization function is a boxcar function, and the spectra are based on an average of about 500 scans over 9 hours.

The sample cell was a multipass White cell with an adjustable absorption path set to  9.6919(23) m (base length is 80 cm). The cell was temperature stabilized by circulating liquid coolant (water) at a temperature of 295.96~K with a combined standard uncertainty of 13~mK. High purity CO (5.5 purity, Linde Gas) was used as supplied. First, the cell was flushed with pure N$_2$ (6.0 purity, Linde Gas) and then a few hPa of pure CO was admitted to the cell, followed by 5 minutes of pumping. Next, pure CO was admitted to the cell, after which the spectrum was measured.  The isotopic 
composition of pure CO sample was determined using a previously measured FTIR spectrum of the (1--0) band. 

Background spectra with an empty pumped cell were measured afterward at a spectral resolution of 0.064 cm$^{-1}$ while the rest of the spectrometer configuration was kept unchanged. Since no optical filter was adopted, no spectral fringes were observed 
in the measured spectra. Later on, we noticed that this choice helped to achieve a smooth rotational $J$-dependence in the retrieved line intensities.

\begin{figure}
\includegraphics[width=0.55\textwidth]{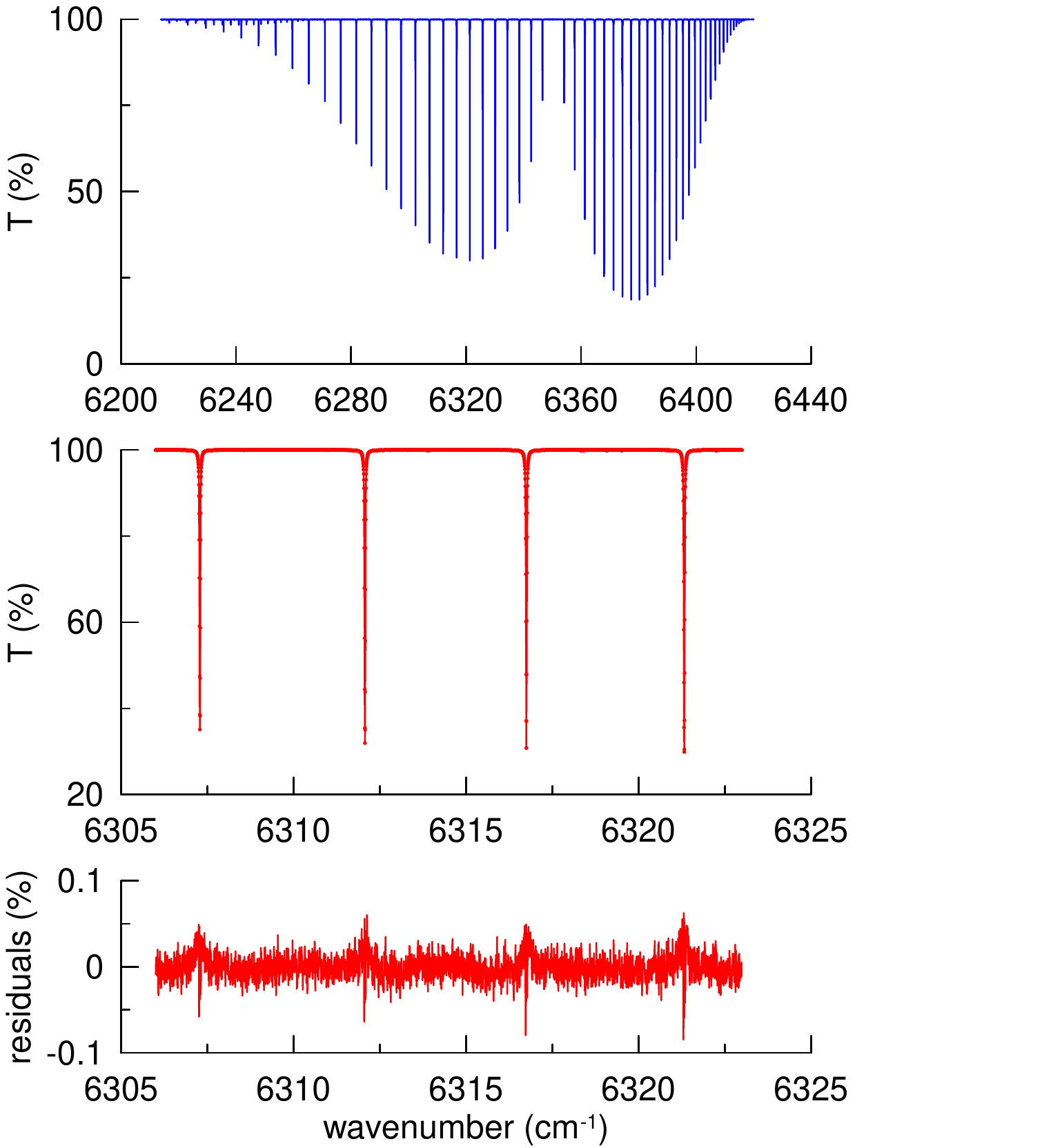}
\caption{Top panel: Normalized transmission spectrum of the CO (3--0) band recorded with a Bruker 125 HR FTIR instrument in the EUMETRISPEC facility at PTB taken at a nominal pressure of 10~kPa and tamperature of 296~K.
Middle panel: Magnified view of the P10, P9, P8 and P7 lines from the spectrum presented in the top panel. 
Bottom panel: Fit residuals (observed - calculated) obtained for the lines shown in the middle panel based on a fit of the speed-dependent Voigt profile (with first-order line-mixing) to the measured spectrum.
For the magnified spectral region the ratio of the maximum peak-to-peak absorption depth to the root-mean-square of the fit residuals is $5100$. }
\label{fig:ptb}
\end{figure}

Spectra were taken with the CO sample at a calibrated pressure of 10.158(7) kPa measured with an MKS Baratron capacitance diaphragm gauge calibrated against a PTB primary pressure standard (0.07 \% relative standard uncertainty).  The 
transmission spectrum $I/I_0$ was fitted with the speed-dependent Voigt function, which is a limiting case of the HTP, with first-order line-mixing coefficients incorporated and fixed to HITRAN 2020 values. The measured spectrum together with an example of the fit residuals is shown in Fig. \ref{fig:ptb}. A third-order polynomial fit of the entire baseline was adopted and the instrumental line shape (ILS) function was precisely determined from separate measurements of N$_2$O lines with the same spectrometer configuration. The fitted line intensities of the P22 to R22 lines scaled to 100 \% 
abundance of $^{12}$C$^{16}$O are presented in Table I. The relatively high accuracy and/or wide dynamic range achieved with this FTS is ascribed to several factors including but not limited to precise temperature stabilization and characterization of the sample path length, operation at a high signal-to-noise ratio (nominally 2000:1), and the use of an InGaAs detector with high linearity.

The averaged relative combined standard uncertainty (approximately 1.3 $\permil$) of the retrieved line intensity is the quadrature summation of pressure uncertainty (0.7 $\permil$), path length uncertainty (0.12 $\permil$), line area uncertainty (0.1 to 0.5 $\permil$), spectrum modeling (1 $\permil$), temperature uncertainty (0.01 to 0.2 $\permil$) and sample isotopic compositions (0.01 $\permil$), sample purity (0.0025 $\permil$). 
The relative precision of the measured line intensities within this band is clearly better than the absolute intensity 
accuracy.  All rotational lines within this band are measured simultaneously due to the nature of Fourier transform spectroscopy. The relative precision only depends on line area uncertainty, temperature effects and relative line shape effect. Combined in a quadrature summation, these terms give a relative precision of 0.6 $\permil$. 

\section{Theoretical calculations}
 \label{sec:calcs}
 
Accuracy of the intensity calculations (both purely \ai\ and semi-empirical) is determined by accuracy of the wave functions and Dipole Moment Curve (DMC). 
Accuracy of the wave functions are based on Potential Energy Curve (PEC) accuracy and on solution of the nuclear Schrodinger equation.

Current ro-vibrational codes for solving nuclear motion problem are extremely accurate. It was shown recently \cite{jtMizus.CO2} that the inaccuracy of  line intensities determined using the code DUO \cite{jt609} for diatomic molecules  is less than 10$^{-5}$ \%. 

So accuracy of the wave functions mainly relies on the accuracy of the PEC. In current work we used the empirical PEC of Coxon {\it et al.} \cite{04Coxon.CO2}, which reproduces the CO energy levels within the experimental uncertainty; this accuracy should be  sufficient to provide accurate wave functions. This the shifts focus of the attention for the accurate intensity calculations to the accuracy of the DMC.

All electronic structure computations were carried out with the quantum chemistry package Molpro\cite{Molpro:JCP:2020}.
Dipoles were computed using the finite differences (FD) approach, which necessitated two calculations per point for the dipoles and one other at zero field to obtain the energy at that geometry.

Dipoles were calculated at the multi-reference configuration
interaction (MRCI) level of theory, the fixed reference
Davidson correction (+Q) has been applied to the MRCI dipoles using aug-cc-pCV6Z  basis set \cite{02PeDuxx.ai}.
Important for an accurate DMC is determination of the complete active space (CAS). The default CAS for CO in  Molpro is (6,2,2,0) in C$_{2v}$ symmetry; use of this CAS is not enough to provide the  desired accuracy for comparison with experiment. 
Intensity calculations on the water molecule showed increasing the CAS
results in significant improvement of the line intensity calculations \cite{jt509}.
In this work  we  chose a CAS with (7,2,2,0)
which gives much better agreement with experiment. 

In general we can try further increases of the CAS to try and further improve the intensity predictions. However, this becomes computationally very expensive for little obvious improvement;  there is no clear dependency between the choice of the CAS and accuracy of the calculations. 
Apart from the CAS, there is a room for further improvement by taking into account different corrections such as  relativistic and adiabatic corrections which are extremely important when we talk about accuracy of  \ai\ PEC, but less crucial when we deal with the DMC.  In case of relativistic correction for intensity calculations we can use the simplest realisation of it - scalar relativistic correction, MVD1, those is produced by Molpro;
the MVD1 correction decreases intensities by about 0.23\% on average. This means that the relativistic correction although important, is minor, and use of  full relativistic Dirac Hamiltonian treatment is unlikely to lead to a significant change.

\section{Comparison of the theory with experimental data of other CO bands}

Here we compare the \ai\ results based on the present calculations for bands other than the (3--0) band with literature data. This comparison includes line intensities for the (1--0), (2--0) and (4--0) bands given in HITRAN 2020. These bands agree with our calculations to within 1.7 \%, 0.1 \% and 2 \%. respectively.  For the (1--0) and (4--0) bands, the HITRAN values are the average of the published experimental data (see \cite{15GangLi.CO2}).
We note that although the (1--0) band is relatively strong and easy to observe, these spectra tend to be optically thick, thus making quantitative determinations of intensities challenging.  This difficulty is manifest in the distribution of observed intensities from six different studies of this band \cite{Hartmann86.CO2, Devi18.CO2, Zou02.CO2, Varanasi75.CO2, Sung05.CO2, Regalia05.CO2}, which exhibit a scatter of 2 \%.  Thus, the most representative comparison to be done in this case is to compare our theoretical calculations with the average values of all
these measurements. The discrepancy of 1.7 \% between our calculations and this average value is comparable to the scatter between various experimental
measurements of this band. For the (2--0) band the agreement between HITRAN 2020, which practically
coincides with the experiment of Devi et al. \cite{Devi12.CO2}, and our  \ai\ calculations is within 0.1 \%.  However, the uncertainties in \cite{Devi12.CO2} did not consider systematic effects. In an attempt to provide more realistic standard uncertainties for these measurements, they were re-evaluated by \cite{15GangLi.CO2} and estimated to range from 0.05 \% to 0.4 \%, although it was not possible to fully account for all important systematic uncertainty components in \cite{Devi12.CO2}. 
 The standard deviation of the relative differences between the intensities of the (4--0) band measured by Li {\it et al.}  \cite{15GangLi.CO2} and our theoretical calculations is about 1.5 \%, which is within the estimated uncertainty of the measurements.  Thus, our
theoretical model allows one to calculate all line intensities in the (4--0) band  within the experimental uncertainty. 
Higher overtones will require additional work because values of the calculated intensities starting from the (5--0) band are affected not only by the accuracy of the quantum chemistry calculations, but also by that of the DMC functional form \cite{Medvedev21.CO2}. 

\section{Using intensities in the (3--0) band of CO as reference intensities}
As mentioned in the main text, the present theoretical (3--0) CO line intensities could be used as intrinsic spectroscopic references to improve measurement accuracy in the case of techniques such as FTS and cavity-enhanced spectroscopy (CEAS) \cite{Gianfrani:99} which require knowledge of the optical path length, $\ell$. This approach would involve measuring the peak area of one or more CO reference lines, $A_{\textrm{CO}}$, which can be modeled as $n_{\textrm{CO}}\ell S_{\textrm{CO}}$, where $n_{\textrm{CO}}$ and $S_{\textrm{CO}}$ are the number density and intensity respectively of the CO line(s). Using the same spectrometer setup, additional measurements of peak areas for an unknown line, $A_u$, would yield the unknown line intensity, $S_u = \left(A_u/A_{\textrm{CO}}\right)\times\left(n_{\textrm{CO}}/n_u\right)$ without an additional measurement of optical path length. With single-component (pure) samples, the ratio $n_{\textrm{CO}}/n_u$, will equal the ratio of $p$/$T$, for both sets of measurements -- a quantity that is likely to have a relatively small uncertainty in most experiments.

\section{Results}\label{results}
Here we present our measured and calculated line intensities for the (3--0) band.
Table \ref{tab:nist_ncu} compares the measurements performed at NCU and NIST and gives weighted averaged line intensities based on these values. Table \ref{tab:obs.-calc.3-0} compares our measured intensities to our  calculated values, and the HITRAN 2020 values are tabulated for reference purposes. 
Analogous theoretical results for (1--0), (2--0) and (4--0) bands are given in Tables \ref{tab:obs.-calc.1-0}, \ref{tab:obs.-calc.2-0} and \ref{tab:obs.-calc.4-0}, respectively.  

\begin{table*}[!ht]
\caption{Comparison of experimental line intensities in the (3--0) of CO. Reported intensities, $S$, are based on the reference temperature $T$ = 296 K, and scaled to 100 \% relative abundance of the $^{12}$C$^{16}$O isotopologue. Intensity units are in cm$^{2}$ cm$^{-1}$/molecule. $S_{\textrm{prev.}}$ indicates previous results obtained at NIST and NCU and $S_{\textrm{av.}}$  is the weighted average from two or three line intensity values given in the table.}
\label{tab:nist_ncu}
\begin{tabular}{cccccc}
\hline\hline
line & wavenumber	&$S_{\textrm{NIST}}$ & $S_{\textrm{NCU}}$ & $S_{\textrm{prev.}}$ &$S_{\textrm{av.}}$   \\
     &  \cm     &  &   &      &             \\
\hline
P30&6190.07&1.5790(28)E-26& 1.5758(18)E-26 &   & 1.5768(16)E-26  \\
P27&6210.25&7.3680(120)E-26& 7.3608(76)E-26& 7.3658(82)E-26$^a$  & 7.3643(51)E-26  \\
R23&6410.88&8.1687(74)E-25& 8.1719(77)E-25 & 8.1659(58)E-25$^b$ & 8.1681(40)E-25 \\
R26&6414.08&2.3661(33)E-25& 2.3668(24)E-25 &  & 2.3665(20)E-25  \\
R27&6414.93&1.5034(21)E-25& 1.5049(16)E-25 &  & 1.5043(13)E-25  \\
R28&6415.67&9.397(18)E-26& 9.386(11)E-26&   & 9.3891(90)E-26  \\
R29&6416.30&5.7300(94)E-26& 5.7371(63)E-26&  & 5.7349(53)E-26  \\

\hline

\multicolumn{6}{l}{$^a$Measurement conducted at NIST in 2018.}\\
\multicolumn{6}{l}{$^b$Measurement conducted at NCU in 2019 \cite{Cygan2019}.}

\end{tabular}
\end{table*}

\newpage
\begin{table*}
\caption{Comparison of experimental line intensities, $S_{\textrm{exp}}$, with 
calculations, $S_{UCL}$, in the (3--0) band of CO. HITRAN 2020 values, 
$S_{\textrm{HT}}$, are given for reference purposes. All intensities are scaled 
to 100 \% relative abundance of the $^{12}$C$^{16}$O isotopologue. Intensity 
units are in cm$^{2}$ cm$^{-1}$/molecule. $S_{\textrm{exp}}$ is either the 
weigted average of values given in Table \ref{tab:nist_ncu} or values determined 
from PTB measurements. }
\label{tab:obs.-calc.3-0}
\begin{tabular}{cccccc}
 \hline\hline
line & wavenumber & $S_{\textrm{HT}}$  & $S_{\textrm{exp}}$ & $S_{\textrm{UCL}}$ & $\left(S_{\textrm{exp}}/S_{\textrm{UCL}}-1\right)$  \\
     &  \cm     & & & & \permil \\
     \hline
P46&6067.26&2.2645E-31& &2.2997E-31&  \\
P45&6075.68&5.2243E-31& &5.3038E-31&  \\
P44&6084.00&1.1839E-30& &1.2009E-30&  \\
P43&6092.22&2.6314E-30& &2.6697E-30&  \\
P42&6100.34&5.7443E-30& &5.8262E-30&  \\
P41&6108.37&1.2316E-29& &1.2481E-29&  \\
P40&6116.29&2.5888E-29& &2.6246E-29&  \\
P39&6124.12&5.3459E-29& &5.4172E-29&  \\
P38&6131.85&1.0826E-28& &1.0974E-28&  \\
P37&6139.47&2.1540E-28& &2.1817E-28&  \\
P36&6147.00&4.2026E-28& &4.2564E-28&  \\
P35&6154.43&8.0483E-28& &8.1484E-28&  \\
P34&6161.76&1.5124E-27& &1.5306E-27&  \\
P33&6168.99&2.7885E-27& &2.8209E-27&  \\
P32&6176.11&5.0418E-27& &5.1004E-27&  \\
P31&6183.14&8.9454E-27& &9.0462E-27&  \\
P30&6190.07&1.5570E-26&1.5768(16)E-26&1.5738E-26&1.9 \\
P29&6196.90&2.6557E-26& &2.6854E-26&  \\
P28&6203.62&4.4458E-26& &4.4936E-26&  \\
P27&6210.25&7.2972E-26&7.3643(51)E-26&7.3735E-26&-1.2 \\
P26&6216.77&1.1738E-25& &1.1863E-25&  \\
P25&6223.19&1.8519E-25& &1.8710E-25&  \\
P24&6229.51&2.8656E-25& &2.8925E-25&  \\
P23&6235.73&4.3424E-25& &4.3825E-25&  \\
P22&6241.85&6.4467E-25&6.4974(85)E-25&6.5066E-25&-1.4 \\
P21&6247.87&9.3792E-25&9.487(13)E-25&9.4640E-25&2.4 \\
P20&6253.78&1.3360E-24&1.3514(18)E-24&1.3483E-24&2.3 \\
P19&6259.59&1.8651E-24&1.8835(25)E-24&1.8810E-24&1.3 \\
P18&6265.30&2.5483E-24&2.5691(34)E-24&2.5690E-24&0.1 \\
P17&6270.91&3.4058E-24&3.4393(45)E-24&3.4335E-24&1.7 \\
P16&6276.42&4.4539E-24&4.4904(59)E-24&4.4893E-24&0.3 \\
P15&6281.82&5.6956E-24&5.7460(75)E-24&5.7393E-24&1.2 \\
P14&6287.12&7.1178E-24&7.1750(94)E-24&7.1706E-24&0.6 \\
P13&6292.32&8.6859E-24&8.759(12)E-24&8.7491E-24&1.2 \\
P12&6297.41&1.0339E-23&1.0426(14)E-23&1.0417E-23&0.9 \\
P11&6302.40&1.2012E-23&1.2112(16)E-23&1.2089E-23&1.9 \\
P10&6307.29&1.3573E-23&1.3675(18)E-23&1.3658E-23&1.2 \\
P9&6312.07&1.4901E-23&1.5017(20)E-23&1.4995E-23&1.5 \\
P8&6316.75&1.5863E-23&1.5982(21)E-23&1.5960E-23&1.4 \\
P7&6321.33&1.6320E-23&1.6431(22)E-23&1.6413E-23&1.1 \\
P6&6325.80&1.6127E-23&1.6251(22)E-23&1.6229E-23&1.4 \\
P5&6330.17&1.5225E-23&1.5321(20)E-23&1.5313E-23&0.5 \\
P4&6334.43&1.3542E-23&1.3615(18)E-23&1.3616E-23&-0.0 \\
P3&6338.59&1.1079E-23&1.1138(15)E-23&1.1140E-23&-0.2 \\
P2&6342.64&7.9115E-24&7.948(11)E-24&7.9519E-24&-0.5 \\
P1&6346.59&4.1579E-24&4.1741(55)E-24&4.1785E-24&-1.1 \\
R0&6354.18&4.3445E-24&4.3617(57)E-24&4.3642E-24&-0.6 \\
R1&6357.81&8.6352E-24&8.670(12)E-24&8.6741E-24&-0.5 \\
R2&6361.34&1.2640E-23&1.2690(17)E-23&1.2691E-23&-0.1 \\
R3&6364.77&1.6127E-23&1.6198(22)E-23&1.6201E-23&-0.2 \\
R4&6368.09&1.8955E-23&1.9039(25)E-23&1.9030E-23&0.5 \\
R5&6371.30&2.0982E-23&2.1075(28)E-23&2.1063E-23&0.6 \\
R6&6374.41&2.2158E-23&2.2262(29)E-23&2.2247E-23&0.7 \\
R7&6377.41&2.2513E-23&2.2609(30)E-23&2.2592E-23&0.7 \\
R8&6380.30&2.2087E-23&2.2189(29)E-23&2.2168E-23&0.9 \\
R9&6383.09&2.1023E-23&2.1099(28)E-23&2.1087E-23&0.6 \\
R10&6385.77&1.9432E-23&1.9505(26)E-23&1.9491E-23&0.7 \\
R11&6388.35&1.7485E-23&1.7551(23)E-23&1.7538E-23&0.8 \\
R12&6390.82&1.5336E-23&1.5390(21)E-23&1.5382E-23&0.5 \\
R13&6393.18&1.3127E-23&1.3173(18)E-23&1.3164E-23&0.7 \\
R14&6395.43&1.0968E-23&1.1009(15)E-23&1.1002E-23&0.7 \\
R15&6397.58&8.9646E-24&8.989(12)E-24&8.9852E-24&0.5 \\
R16&6399.62&7.1593E-24&7.1773(94)E-24&7.1752E-24&0.3 \\
R17&6401.55&5.5943E-24&5.6141(74)E-24&5.6050E-24&1.6 \\
R18&6403.38&4.2766E-24&4.2883(56)E-24&4.2847E-24&0.8 \\
R19&6405.09&3.2011E-24&3.2139(42)E-24&3.2063E-24&2.4 \\
R20&6406.70&2.3456E-24&2.3469(31)E-24&2.3494E-24&-1.1 \\
R21&6408.20&1.6837E-24&1.6864(22)E-24&1.6861E-24&0.2 \\
R22&6409.60&1.1839E-24&1.1876(16)E-24&1.1854E-24&1.8 \\
R23&6410.88&8.1568E-25&8.1681(40)E-25&8.1663E-25&0.2 \\
R24&6412.06&5.5071E-25& &5.5131E-25&  \\
R25&6413.12&3.6440E-25& &3.6481E-25&  \\
R26&6414.08&2.3638E-25&2.3665(20)E-25&2.3663E-25&0.1 \\
R27&6414.93&1.5042E-25&1.5043(13)E-25&1.5049E-25&-0.4 \\
R28&6415.67&9.3782E-26&9.3891(90)E-26&9.3839E-26&0.6 \\
R29&6416.30&5.7352E-26&5.7349(53)E-26&5.7381E-26&-0.6 \\
R30&6416.82&3.4403E-26& &3.4411E-26&  \\
R31&6417.24&2.0242E-26& &2.0240E-26&  \\
R32&6417.54&1.1677E-26& &1.1677E-26&  \\
R33&6417.73&6.6089E-27& &6.6090E-27&  \\
R34&6417.81&3.6704E-27& &3.6696E-27&  \\
R35&6417.79&1.9999E-27& &1.9991E-27&  \\
R36&6417.65&1.0694E-27& &1.0685E-27&  \\
R37&6417.40&5.6064E-28& &5.6044E-28&  \\
R38&6417.04&2.8858E-28& &2.8845E-28&  \\
R39&6416.57&1.4576E-28& &1.4570E-28&  \\
R40&6415.99&7.2283E-29& &7.2227E-29&  \\
R41&6415.30&3.5173E-29& &3.5143E-29&  \\
R42&6414.50&1.6796E-29& &1.6783E-29&  \\
R43&6413.58&7.8760E-30& &7.8679E-30&  \\
R44&6412.56&3.6248E-30& &3.6207E-30&  \\
R45&6411.42&1.6380E-30& &1.6358E-30&  \\
R46&6410.17&7.2648E-31& &7.2551E-31&  \\
R47&6408.81&3.1636E-31& &3.1593E-31&  \\
R48&6407.34&1.3532E-31& &1.3508E-31&  \\
\hline
\end{tabular}
\end{table*}
\newpage

\begin{table*}
\caption{Calculated intensities of (1--0) band.  Reported intensities, $S$, are based on the reference temperature $T$ = 296 K, and scaled to 100 \% relative abundance of the $^{12}$C$^{16}$O isotopologue. Intensity units are in cm$^{2}$ cm$^{-1}$/molecule.}
\label{tab:obs.-calc.1-0}

\begin{tabular}{ccc}
\hline
line & wavenumber	&  $S_{\textrm{UCL}}$  \\
     \hline
P40&1963.729&8.416E-25  \\
P39&1968.825&1.723E-24  \\
P38&1973.891&3.464E-24  \\
P37&1978.929&6.832E-24  \\
P36&1983.936&1.322E-23  \\
P35&1988.914&2.510E-23  \\
P34&1993.862&4.677E-23  \\
P33&1998.780&8.548E-23  \\
P32&2003.668&1.533E-22  \\
P31&2008.525&2.695E-22  \\
P30&2013.352&4.649E-22 \\
P29&2018.149&7.865E-22  \\
P28&2022.915&1.305E-21  \\
P27&2027.649&2.122E-21 \\
P26&2032.353&3.384E-21  \\
P25&2037.025&5.290E-21  \\
P24&2041.667&8.105E-21  \\
P23&2046.276&1.217E-20  \\
P22&2050.854&1.790E-20 \\
P21&2055.401&2.580E-20 \\
P20&2059.915&3.642E-20 \\
P19&2064.397&5.034E-20 \\
P18&2068.847&6.811E-20 \\
P17&2073.265&9.018E-20 \\
P16&2077.650&1.168E-19 \\
P15&2082.002&1.479E-19 \\
P14&2086.322&1.830E-19 \\
P13&2090.609&2.211E-19 \\
P12&2094.862&2.607E-19 \\
P11&2099.083&2.996E-19 \\
P10&2103.270&3.352E-19 \\
P9 &2107.423&3.644E-19 \\
P8 &2111.543&3.840E-19 \\
P7 &2115.629&3.909E-19 \\
P6 &2119.681&3.827E-19 \\
P5 &2123.699&3.575E-19 \\
P4 &2127.683&3.146E-19 \\
P3 &2131.632&2.548E-19 \\
P2 &2135.546&1.800E-19 \\
P1 &2139.426&9.362E-20 \\
R0 &2147.081&9.577E-20 \\
R1 &2150.856&1.884E-19 \\
R2 &2154.596&2.727E-19 \\
R3 &2158.300&3.445E-19 \\
R4 &2161.968&4.004E-19 \\
R5 &2165.601&4.384E-19 \\
R6 &2169.198&4.582E-19 \\
R7 &2172.759&4.603E-19 \\
R8 &2176.284&4.468E-19 \\
R9 &2179.772&4.205E-19 \\
R10&2183.224&3.845E-19 \\
R11&2186.639&3.422E-19 \\
R12&2190.018&2.968E-19 \\
R13&2193.359&2.513E-19 \\
R14&2196.664&2.077E-19 \\
R15&2199.931&1.678E-19 \\
R16&2203.161&1.325E-19 \\
R17&2206.354&1.024E-19 \\
R18&2209.508&7.737E-20 \\
R19&2212.626&5.725E-20 \\
R20&2215.705&4.148E-20 \\
R21&2218.746&2.944E-20 \\
R22&2221.748&2.046E-20 \\
R23&2224.713&1.394E-20 \\
R24&2227.639&9.303E-21 \\
R25&2230.526&6.086E-21 \\
R26&2233.374&3.903E-21 \\
R27&2236.184&2.454E-21 \\
R28&2238.954&1.512E-21 \\
R29&2241.685&9.142E-22 \\
R30&2244.377&5.419E-22  \\
R31&2247.029&3.151E-22  \\
R32&2249.641&1.797E-22  \\
R33&2252.214&1.005E-22  \\
R34&2254.747&5.516E-23  \\
R35&2257.239&2.970E-23  \\
R36&2259.691&1.569E-23  \\
R37&2262.103&8.131E-24  \\
R38&2264.474&4.136E-24  \\
R39&2266.805&2.064E-24  \\
R40&2269.096&1.011E-24  \\

\hline
 
\end{tabular}
\end{table*}

\newpage
\begin{table*}
\caption{Calculated intensities of (2--0) band. Reported intensities, $S$, are based on the reference temperature $T$ = 296 K, and scaled to 100 \% relative abundance of the $^{12}$C$^{16}$O isotopologue. Intensity units are in cm$^{2}$ cm$^{-1}$/molecule.}
\label{tab:obs.-calc.2-0}

\begin{tabular}{ccc}
\hline
line & wavenumber	&  $S_{\textrm{UCL}}$  \\
     \hline
P40&4053.217&5.437E-27  \\
P39&4059.679&1.116E-26  \\
P38&4066.075&2.250E-26  \\
P37&4072.408&4.449E-26  \\
P36&4078.675&8.634E-26  \\
P35&4084.878&1.644E-25  \\
P34&4091.017&3.072E-25  \\
P33&4097.090&5.630E-25  \\
P32&4103.098&1.012E-24  \\
P31&4109.040&1.786E-24  \\
P30&4114.917&3.089E-24 \\
P29&4120.729&5.242E-24  \\
P28&4126.474&8.722E-24  \\
P27&4132.154&1.423E-23 \\
P26&4137.768&2.276E-23  \\
P25&4143.316&3.570E-23  \\
P24&4148.797&5.487E-23  \\
P23&4154.211&8.266E-23  \\
P22&4159.560&1.220E-22 \\
P21&4164.841&1.764E-22 \\
P20&4170.055&2.498E-22 \\
P19&4175.202&3.465E-22 \\
P18&4180.283&4.704E-22 \\
P17&4185.295&6.250E-22 \\
P16&4190.240&8.122E-22 \\
P15&4195.118&1.032E-21 \\
P14&4199.928&1.282E-21 \\
P13&4204.670&1.554E-21 \\
P12&4209.343&1.839E-21 \\
P11&4213.949&2.122E-21 \\
P10&4218.486&2.382E-21 \\
P9 &4222.954&2.599E-21 \\
P8 &4227.354&2.749E-21 \\
P7 &4231.685&2.810E-21 \\
P6 &4235.947&2.761E-21 \\
P5 &4240.140&2.589E-21 \\
P4 &4244.264&2.288E-21 \\
P3 &4248.318&1.860E-21 \\
P2 &4252.302&1.319E-21 \\
P1 &4256.217&6.889E-22 \\
R0 &4263.837&7.104E-22 \\
R1 &4267.542&1.403E-21 \\
R2 &4271.177&2.040E-21 \\
R3 &4274.741&2.587E-21 \\
R4 &4278.234&3.019E-21 \\
R5 &4281.657&3.320E-21 \\
R6 &4285.009&3.484E-21 \\
R7 &4288.290&3.516E-21 \\
R8 &4291.499&3.427E-21 \\
R9 &4294.638&3.239E-21 \\
R10&4297.705&2.974E-21 \\
R11&4300.700&2.659E-21 \\
R12&4303.623&2.317E-21 \\
R13&4306.475&1.970E-21 \\
R14&4309.254&1.636E-21 \\
R15&4311.962&1.327E-21 \\
R16&4314.597&1.053E-21 \\
R17&4317.159&8.170E-22 \\
R18&4319.649&6.204E-22 \\
R19&4322.066&4.612E-22 \\
R20&4324.410&3.357E-22 \\
R21&4326.681&2.394E-22 \\
R22&4328.879&1.672E-22 \\
R23&4331.003&1.144E-22 \\
R24&4333.054&7.672E-23 \\
R25&4335.031&5.043E-23 \\
R26&4336.934&3.250E-23 \\
R27&4338.764&2.053E-23 \\
R28&4340.519&1.272E-23 \\
R29&4342.200&7.724E-24 \\
R30&4343.807&4.601E-24  \\
R31&4345.339&2.688E-24  \\
R32&4346.796&1.541E-24  \\
R33&4348.179&8.661E-25  \\
R34&4349.486&4.777E-25  \\
R35&4350.719&2.585E-25  \\
R36&4351.876&1.372E-25  \\
R37&4352.958&7.149E-26  \\
R38&4353.964&3.655E-26  \\
R39&4354.895&1.834E-26  \\
R40&4355.749&9.029E-27  \\
\hline

\end{tabular}
\end{table*}

\newpage
\begin{table*}
\caption{Calculated intensities of (4--0) band. Reported intensities, $S$, are based on the reference temperature $T$ = 296 K, and scaled to 100 \% relative abundance of the $^{12}$C$^{16}$O isotopologue. Intensity units are in cm$^{2}$ cm$^{-1}$/molecule.}
\label{tab:obs.-calc.4-0}

\begin{tabular}{ccc}
\hline
line & wavenumber	&  $S_{\textrm{UCL}}$  \\
     \hline
P38&8171.272&1.265E-31  \\
P37&8180.195&2.610E-31  \\
P36&8188.983&5.282E-31  \\
P35&8197.636&1.048E-30  \\
P34&8206.154&2.039E-30  \\
P33&8214.537&3.890E-30  \\
P32&8222.785&7.278E-30  \\
P31&8230.898&1.335E-29  \\
P30&8238.875&2.400E-29 \\
P29&8246.716&4.230E-29  \\
P28&8254.422&7.308E-29  \\
P27&8261.992&1.237E-28 \\
P26&8269.425&2.053E-28  \\
P25&8276.723&3.338E-28  \\
P24&8283.884&5.318E-28  \\
P23&8290.909&8.299E-28  \\
P22&8297.797&1.268E-27 \\
P21&8304.548&1.898E-27 \\
P20&8311.163&2.782E-27 \\
P19&8317.640&3.989E-27 \\
P18&8323.980&5.599E-27 \\
P17&8330.183&7.687E-27 \\
P16&8336.248&1.032E-26 \\
P15&8342.175&1.354E-26 \\
P14&8347.965&1.736E-26 \\
P13&8353.617&2.172E-26 \\
P12&8359.130&2.652E-26 \\
P11&8364.506&3.154E-26 \\
P10&8369.743&3.651E-26 \\
P9 &8374.841&4.105E-26 \\
P8 &8379.801&4.473E-26 \\
P7 &8384.622&4.707E-26 \\
P6 &8389.304&4.762E-26 \\
P5 &8393.847&4.596E-26 \\
P4 &8398.251&4.178E-26 \\
P3 &8402.515&3.494E-26 \\
P2 &8406.639&2.549E-26 \\
P1 &8410.624&1.368E-26 \\
R0 &8418.174&1.490E-26 \\
R1 &8421.739&3.023E-26 \\
R2 &8425.164&4.513E-26 \\
R3 &8428.448&5.877E-26 \\
R4 &8431.591&7.040E-26 \\
R5 &8434.594&7.945E-26 \\
R6 &8437.456&8.554E-26 \\
R7 &8440.177&8.853E-26 \\
R8 &8442.757&8.851E-26 \\
R9 &8445.195&8.576E-26 \\
R10&8447.492&8.073E-26 \\
R11&8449.647&7.396E-26 \\
R12&8451.661&6.603E-26 \\
R13&8453.532&5.751E-26 \\
R14&8455.262&4.891E-26 \\
R15&8456.849&4.064E-26 \\
R16&8458.294&3.301E-26 \\
R17&8459.596&2.622E-26 \\
R18&8460.756&2.038E-26 \\
R19&8461.773&1.550E-26 \\
R20&8462.647&1.155E-26 \\
R21&8463.378&8.420E-27 \\
R22&8463.966&6.014E-27 \\
R23&8464.410&4.209E-27 \\
R24&8464.711&2.886E-27 \\
R25&8464.868&1.939E-27 \\
R26&8464.882&1.277E-27 \\
R27&8464.751&8.245E-28 \\
R28&8464.476&5.219E-28 \\
R29&8464.057&3.238E-28 \\
R30&8463.494&1.971E-28  \\
R31&8462.786&1.176E-28  \\
R32&8461.934&6.883E-29  \\
R33&8460.936&3.951E-29  \\
R34&8459.794&2.225E-29  \\
R35&8458.506&1.229E-29  \\
R36&8457.073&6.661E-30  \\
R37&8455.495&3.542E-30  \\
R38&8453.771&1.848E-30  \\
R39&8451.901&9.459E-31  \\
R40&8449.886&4.752E-31  \\

\hline
 
\end{tabular}
\end{table*}

\bibliography{journals_phys,jtj,programs,CO2,ncu_nist,N2O}

\end{document}